\documentclass[]{pasj02} 
\usepackage[switch,mathlines]{lineno} 
\usepackage{natbib} 

\jyear{2024}
\Received{2024/08/30}
\Accepted{2024/10/29}

\usepackage{newtxtext,newtxmath}
\usepackage{color}
\usepackage{physics}
\newcommand{\vmax}{V_\mathrm{max}}
\newcommand{\rmax}{r_\mathrm{max}}

\begin{document} 

\title{
Cosmological evolution of dark matter subhaloes under tidal stripping by growing Milky Way--like galaxies
}


\author{
 Yudai \textsc{Kazuno}\altaffilmark{1}$^{\dag}$\orcid{N/A},
 Masao \textsc{Mori}\altaffilmark{2}\altemailmark\orcid{0000-0002-2883-8943} \email{mmori@ccs.tsukuba.ac.jp},
 Yuka \textsc{Kaneda}\altaffilmark{1}\orcid{0009-0002-3547-9472},
 and
 Koki \textsc{Otaki}\altaffilmark{3, 4}\orcid{0000-0002-3406-3099}
}

\altaffiltext{1}{Graduate School of Science and Technology, University of Tsukuba, 1-1-1 Tennodai, Tsukuba, 305-8577 Ibaraki, Japan}
\altaffiltext{2}{Center for Computational Sciences, University of Tsukuba, 1-1-1 Tennodai, Tsukuba, 305-8577 Ibaraki, Japan}
\altaffiltext{3}{Amanogawa Galaxy Astronomy Research Center, Graduate School of Science and Engineering, Kagoshima University, 1-21-35 Korimoto, Kagoshima, 390-0065 Kagoshima, Japan}
\altaffiltext{4}{Dipartimento di Fisica, Sapienza Universita di Roma, Piazzale Aldo Moro 5, I-00185 Rome, Italy}
\altaffiltext{$\dag$}{Present address: Urawa Girls' Upper Secondary School, 3-8-45 Kishicho, Urawa-ku, Saitama City, 330-0064 Saitama, Japan}



\maketitle


\begin{abstract}
We present the findings of a comprehensive and detailed analysis of merger tree data from ultra--high--resolution cosmological $N$--body simulations. The analysis, conducted with a particle mass resolution of $5 \times 10^3 h^{-1} M_{\odot}$ and a halo mass resolution of $10^7 h^{-1} M_{\odot}$, provides sufficient accuracy to suppress numerical artefacts.
This study elucidates the dynamical evolution of subhaloes associated with the Milky Way--like host haloes. Unlike more massive dark matter haloes, which have been extensively studied, these subhaloes follow a distinct mass evolution pattern: an initial accretion phase, followed by a tidal stripping phase where mass is lost due to the tidal forces of the host halo.
The transition from accretion to stripping, where subhaloes reach their maximum mass, occurs around a redshift of $z\simeq1$. Smaller subhaloes reach this point earlier, while larger ones do so later. Our analysis reveals that over 80 per cent of subhaloes have experienced mass loss, underscoring the universality of tidal stripping in subhalo evolution.
Additionally, we derived the eccentricities and pericentre distances of subhalo orbits from the simulations and compare them with those of nearby satellite galaxies observed by the Gaia satellite. The results demonstrate a significant alignment between the orbital elements predicted by the cold dark matter model and the observed data, providing robust support for the model as a credible candidate for dark matter.
\end{abstract}





\section{Introduction}
Understanding the physical nature of dark matter, which constitutes the majority of the matter density in the Universe, has long been a central objective in both physics and cosmology. The widely accepted model posits that dark matter is mostly composed of cold dark matter (CDM). While the CDM model successfully accounts for the large--scale structure of galaxy distribution, it encounters serious discrepancies between theory and observation  on subgalactic scales including the cusp--core problem \citep{Moore1994, Burkert1995}, the too--big--to--fail problem \citep{Boylan-KolchinBullockKaplinghat2011}, and the missing satellite problem \citep{Moore+1999, Klypin+1999}.

Within the CDM framework, the most promising resolution to these discrepancies involves stellar feedback mechanisms that alter the gravitational potential of the systems \citep{NavarroFrenkWhite1996, GnedinZhao2002, ReadGilmore2005, OgiyaMori2014, BullockBoylan-Kolchin2017}. Most recently, \citet{KanedaMoriOtaki2024} conducted a comprehensive analysis of dark matter halo scaling relations and the cusp--core problem within the CDM framework, encompassing systems ranging from dwarf galaxies to galaxy clusters. Their findings underscore the critical role that the mass distribution and the evolutionary history of subgalactic dark matter haloes play in elucidating the true nature of dark matter.

Given these complexities, elucidating the growth of dark matter haloes is fundamentally important for advancing physical cosmology. Significant effort has been dedicated to developing analytical models describing the mass acquisition history of dark matter haloes, most of which are derived from merger trees representing the hierarchical merging of dark matter haloes. Such merger trees have been employed in a wide range of cosmological simulations \citep{KauffmannWhite1993, Wechsler+2002, NeisteinDekel2008, McBrideFakhouriMa2009, Zhao+2009, FakhouriMaBoylan-Kolchin2010, Genel+2010, GiocoliTormenSheth2012, Ludlow+2013} as well as statistical models \citep{KauffmannWhite1993, SomervilleKolatt1999, vandenBosch2002, Ogiya+2019}. These models have proven effective in predicting the mass accretion history of dark matter haloes across a wide range of masses and redshifts.

In the hierarchical structure formation scenario posited by the CDM universe, dark matter haloes form from density fluctuations in the early universe. In their exploration of the mass accretion history of dark matter haloes with masses ranging from $10^{12}$ to $10^{14} M_{\odot}$, \citet{Wechsler+2002} demonstrated that the following exponential function could universally fit the mass evolution of dark matter haloes:
\begin{equation}
M(z)=M_0\exp(-\alpha z),
\label{eq:1-param}
\end{equation}
where $M(z)$ is the mass of a dark matter halo at redshift $z$, $M_0$ is its mass at $z=0$, and $\alpha$ is a free parameter determining the growth rate. This model, along with further efforts \citep{vandenBosch+2014}, has shown good agreement for haloes in the mass range from $10^{11}h^{-1} M_\odot$ to $10^{14} h^{-1} M_\odot$ in simulations such as the Bolshoi Simulation \citep{KlypinTrujillo-GomezPrimack2011}, particularly up to $z=3$. However, alternative studies have proposed that the mass evolution of dark matter haloes might also be explained by a power law function \citep{vandenBosch2002}. Furthermore, \citet{Tasitsiomi+2004} and \citet{McBrideFakhouriMa2009} suggested a combination of exponential and power law functions, such as:
\begin{equation}
M(z)=M_0(1+z)^{\beta}\exp(-\gamma z),
\label{eq:2-param}
\end{equation}
where $\beta\lesseqgtr0,\gamma\geq0$ are free parameters. Specifically, \citet{McBrideFakhouriMa2009} demonstrated that the dynamical growth of 500,000 dark matter haloes in the Millennium Simulation \citep{Springel+2005} could be categorised into four patterns based on the parameters $\beta$ and $\gamma$ derived from fitting equation \eqref{eq:2-param}. A similar analysis conducted by \citet{FakhouriMaBoylan-Kolchin2010} using the Millennium--II Simulation supported these findings, illustrating that smaller dark matter haloes evolve into larger ones through repeated mergers and collisions. 

Thus far, studies have primarily focused on haloes linked to massive galaxies and clusters, while the mass evolution of less--massive galaxies associated with Milky Way--sized galaxies in a cosmological context remains ongoing.
\citet{DiemandKuhlenMadau2007} highlighted the role of accretion and tidal stripping in galaxy evolution, though their study focused on just one Milky Way--sized galaxy. In contrast, \citet{Morinaga+2019} asserted, based on cosmological $N$--body simulations, that most subhaloes in nine Milky Way--sized haloes at $z=0$ were accreted between $0.5\leq z\leq 2.5$, while those accreted at higher redshifts suffered complete tidal disruption. 
However, due to computational limitations, further investigation is required into the mass evolution histories of low--mass dark matter haloes on cosmological scales and the dependence of the tidal stripping process on orbital dynamics.

In this study, we analyse the ultra--high--resolution cosmological $N$--body simulation to investigate the growth of subhaloes associated with 27 Milky Way--sized haloes and 1,652,109 subhaloes, achieving better numerical resolution and statistical reliability than previous studies.
We also leverage the recent observations from the Gaia space observatory, which provide high--precision measurements of the orbital motion of Milky Way satellites. By directly comparing the statistical properties of subhalo orbital motion predicted by CDM theory with the actual orbital motions of satellite galaxies, particularly their pericentric distances and eccentricities, we aim to deepen our understanding of these processes.
This paper is organised as follows: Section 2 describes the data set of the cosmological $N$--body simulation and outlines the selection of host and subhaloes, along with the construction of their mass evolutionary histories. Section 3 presents these histories, discusses the impact of tidal forces, and compares the results with recent observational data. Finally, Section 4 summarises our conclusions.
Throughout this paper, we assume cosmological parameters adopted by \citet{Ishiyama+2021}.

\begin{figure*}[t]
    \begin{center}
        \includegraphics[width=\linewidth]{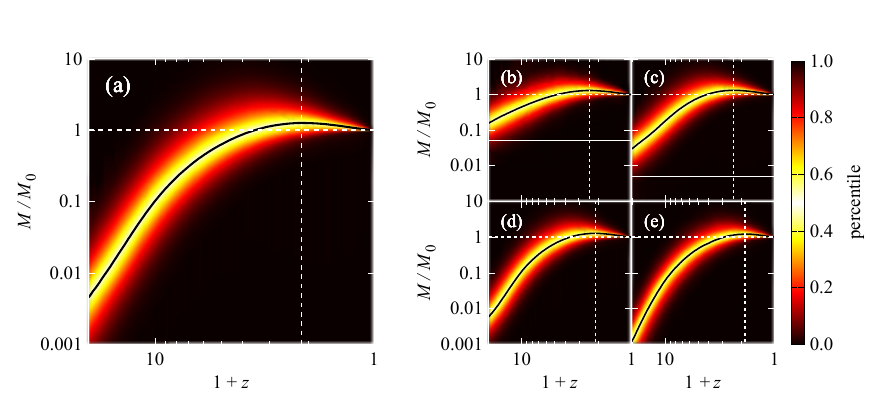}
    \end{center}
    \caption{The mass evolution of the most massive progenitors of subhaloes as a function of redshift. Each panel presents the result for the most massive progenitor of subhaloes, normalised by their final mass $M_0$, where $M_0$ is (a) $M_0\geq 10^7$, (b) $10^5\leq M_0<10^6$, (c) $10^6\leq M_0<10^7$, (d) $10^7\leq M_0<10^8$, and (e) $M_0\geq 10^8$ in units of $h^{-1}M_{\odot}$, respectively. The black solid curves, the horizontal white dashed lines, and the vertical white dashed lines correspond to the median value at a given redshift, $M/M_0=1$, and the peak redshift $z_{\mathrm{f}}$ when the median reaches its maximum value, respectively. The limit of particle mass resolution in \textsc{Phi--4096} is indicated by the white solid line in panel (b) for the mass of $M_0=10^5\ h^{-1}M_{\odot}$ and in panel (c) for the mass of $M_0=10^6\ h^{-1}M_{\odot}$. The colour scale represents the percentile at each output redshift.}
    \label{fig:mass_evolve}
\end{figure*}


\section{Simulation datasets and methods of analysis}
In this study, we examine the dynamical evolution of dark matter subhaloes that are gravitationally bound to Milky Way--like host haloes. In this section, we describe the properties of the data set derived from the cosmological $N$--body simulation and explain how we construct mass evolutionary histories for selected subhaloes.

We analyse data from an ultra--high--resolution cosmological $N$--body simulation, \textsc{Phi--4096} \citep{Ishiyama+2021}. This simulation was employed to assess the impact of mass resolution within the state--of--the--art Uchuu suite and to investigate the mass--concentration relation for dark matter haloes above $10^7\ h^{-1}M_{\odot}$ at high redshift ($z>7.5$). The data set consists of $4096^3$ dark matter particles within a comoving box with a side length of $16.0\ h^{-1}\mathrm{Mpc}$, and a particle mass of $5.13 \times 10^3\ h^{-1}M_{\odot}$. The gravitational softening length is $0.06\ h^{-1}\rm{kpc}$.
To identify gravitationally bound haloes and subhaloes within the particle data at each snapshot, the \textsc{rockstar} phase--space halo/subhalo finder was applied \citep{BehrooziWechslerWu2013}. 
The density profile of each dark matter halo was fitted using the NFW profile proposed by \citet{NavarroFrenkWhite1996}:
\begin{equation}
\rho_{\mathrm{NFW}}(r)=\frac{\rho_{\mathrm{s}}}{(c_{\mathrm{vir}} r/r_{\mathrm{vir}})(1+c_{\mathrm{vir}} r/r_{\mathrm{vir}})^2},
\end{equation}
where $r$ is a distance from the centre, $r_{\mathrm{vir}}$ is a virial radius, $c_{\mathrm{vir}}$ is the concentration parameter and $\rho_{\mathrm{s}}$ is a scale density, respectively. 
To construct the halo and subhalo catalogues and their merger trees, the \textsc{consistent trees} \citep{Behroozi+2013} code was utilised. These data are publicly available on the Skies \& Universes platform (https://skiesanduniverses.org). 
A total of 151 snapshots were taken from \textsc{Phi--4096}, spanning from $z=30$ to the present.

Initially, a total of 27 host haloes were identified within the mass range:
\begin{equation}
3.40\times 10^{11}\ h^{-1}M_{\odot}<M_0<2.04\times 10^{12}\ h^{-1}M_{\odot},
\end{equation}
which broadly encompasses Milky Way--like galaxies \citep[for recent review,][]{BobylevBaykova2023}.
To include subhaloes potentially bound to the host halo, we identified those within twice the virial radius of the host halo at $z=0$, and extracted their corresponding merger tree data.
This selection yielded 32,444 subhaloes with $M_0\geq 10^7\ h^{-1}M_{\odot}$ and 1,652,109 subhaloes with $M_0\geq 10^5\ h^{-1}M_{\odot}$, associated with the 27 host haloes.
It should be noted that subhaloes with masses less than $10^7 h^{-1}M_\odot$ at $z=0$ have artificially short relaxation times of under 10 Gyr, owing to the limited number of particles. Consequently, the effect of tidal stripping may be somewhat overestimated for these less massive subhaloes.

Using the merger tree catalogue, we constructed the mass evolutionary histories for the selected subhaloes. The merger tree catalogue records the merging history of dark matter haloes and is typically represented as a relationship between a "progenitor" dark matter halo and its "descendant" halo. Conventionally, the descendant is linked to the most massive progenitor at the previous timestep. In the \textsc{consistent trees} code, the most massive progenitor is identified as the most massive dark matter halo among "sibling" haloes (excluding "full cousin" haloes) at a given time. We then iteratively traced the progenitors of these most massive progenitors to determine the main evolutionary branch for each subhalo. This approach enabled us to complete the mass evolutionary histories.

We then applied equation \eqref{eq:2-param} to fit the mass evolutionary histories, classifying the evolution types according to the two--parameter model by \citet{McBrideFakhouriMa2009} involving $\beta$ and $\gamma$. The details of this classification are discussed in Section 3.2. Furthermore, we derived the concentration parameter $c_{\mathrm{vir}}$, the formation redshift $z_{\mathrm{f}}$, which indicates the rate of dynamical evolution of subhaloes, and the orbital parameters from these mass evolutionary histories.


\section{Results of analysis}


\subsection{Mass evolutionary histories of subhaloes}

In figure \ref{fig:mass_evolve}, we present the mass evolution of the most massive progenitors of subhaloes, normalised by their virial mass at $z=0$, as a function of redshift. The black solid curve in each panel represents the median value at the given redshift. The colour scale indicates the percentile at each redshift. The horizontal white dashed line shows $M/M_0=1$, and the vertical white dashed line corresponds to the redshift at which the median reaches its maximum value. 

Panel (a) displays the results for the most massive progenitors of subhaloes with $M_0\geq 10^7\ h^{-1}M_{\odot}$. It is evident that $M/M_0$ exceeds unity at a characteristic redshift in the past, indicating that the mass of subhaloes increases at higher redshifts and subsequently decreases at lower redshifts.
This behaviour differs significantly from the dynamical evolutionary path shown in previous studies for isolated dark matter haloes with $M_0>10^{12}\ h^{-1} M_{\odot}$, which only increase in mass over time \citep{Wechsler+2002, vandenBosch2002}. In contrast, this trend is consistent with previous studies on the dynamical evolution of less massive satellite subhaloes \citep{DiemandKuhlenMadau2007}.

Here, we define the peak redshift $z_{\mathrm{peak}}$ as the epoch at which the median (indicated by the vertical white dashed line) in the evolution of subhaloes reaches its maximum. The peak redshift in panel (a) is $z_{\mathrm{peak}}=1.146$. 
It should be emphasised that a significant fraction of all subhaloes with $M_0\geq 10^7\ h^{-1}M_{\odot}$ undergo a mass--loss phase during their evolution in our analysis.
In other words, the mass--loss phase is a universal process in the dynamical evolution of subhaloes.

The dynamical evolutionary history of subhaloes reveals two distinct phases. During the accretion phase, the total mass of subhaloes increases due to gravitational collapse and the merger of subhaloes. After the peak redshift $z_{\mathrm{peak}}$, the total mass of subhaloes decreases due to the tidal force exerted by the host halo in the tidal stripping phase. 
At redshifts $z\gtrsim1$, dark matter haloes acquire mass through the accretion of surrounding smoothly distributed dark matter and more less--massive dark matter haloes. 
As the host halo grows, the effect of tidal forces on subhaloes gradually increases over time. When the tidal force from the host halo exceeds the self--gravity of subhaloes, dark matter particles in the outskirts of subhaloes are easily stripped away.

The four panels on the right in figure \ref{fig:mass_evolve} depict the mass evolution of the most massive progenitors of subhaloes, categorised by their final mass $M_0$. The mass ranges are as follows: (b) $10^5\leq M_0<10^6$, (c) $10^6\leq M_0<10^7$, (d) $10^7\leq M_0<10^8$, and (e) $M_0\geq 10^8$ in units of $h^{-1}M_{\odot}$. It is apparent that almost all subhaloes with $M_0\geq 10^5\ h^{-1}M_{\odot}$ exhibit the tidal stripping phase in their dynamical evolutionary histories.
The limit of mass resolution in \textsc{Phi--4096} is represented by the white solid line in panel (b) for subhaloes with a mass of $M_0=10^5\ h^{-1}M_{\odot}$ and in panel (c) for subhaloes with a mass of $M_0=10^6\ h^{-1}M_{\odot}$. There is significant ambiguity below these lines.
The peak redshifts for the different mass ranges are (b) $z_{\mathrm{peak}}=1.408$, (c) $z_{\mathrm{peak}}=1.353$, (d) $z_{\mathrm{peak}}=1.146$, and (e) $z_{\mathrm{peak}}=0.869$, respectively. 
Due to the counteraction between mass accretion and tidal stripping by the growing host halo, $z_\mathrm{peak}$ is established, with subhaloes reaching their maximum mass at approximately $z_\mathrm{peak} \sim 1$. Additionally, we observed a negative correlation between $z_\mathrm{peak}$ and the mass of the subhaloes.


\subsection{Classification of the mass evolutionary history}

As mentioned earlier, \citet{McBrideFakhouriMa2009} classified the dynamical evolutionary histories of dark matter haloes into four types based on the two parameters in equation \eqref{eq:2-param}. Table \ref{tab:McBride_type} summarises the criteria and characteristics of these four categorised evolutionary paths of dark matter haloes. Type I haloes have a weak contribution from the $(1 + z)$ term and show only slight deviations from an exponential curve. Haloes with $\beta-\gamma<-0.45$ are classified as Type II and exhibit steep growth at late times, while Type III haloes, with fit parameters between $-0.45<\beta-\gamma<0$, show flat late--time growth. Type IV haloes, with $\beta-\gamma>0$, represent the most extreme deviation from an exponential.

The rate of change of mass in equation \eqref{eq:2-param} is represented by
\begin{equation}
\frac{d\ln{M(z)}}{dz}=\frac{\beta}{1+z}-\gamma \approx \beta-\gamma+\order{z}
\end{equation}
at low redshifts. Therefore, the index $\beta-\gamma$ in Table \ref{tab:McBride_type} is appropriate for categorising the evolutionary history of dark matter haloes. \citet{McBrideFakhouriMa2009} also reported that the "average" evolutionary history of isolated dark matter haloes with $M_0>10^{12}\ h^{-1} M_{\odot}$ is approximately described by the exponential function, whereas the path of dynamical evolution significantly deviates from exponential growth for individual dark matter haloes. They asserted that there is a mass dependence on the evolving types and that about 90\% of all dark matter haloes belong to Types I, II, and III, which only increase in mass.

\begin{figure}
    \begin{center}
        \includegraphics[width=\columnwidth]{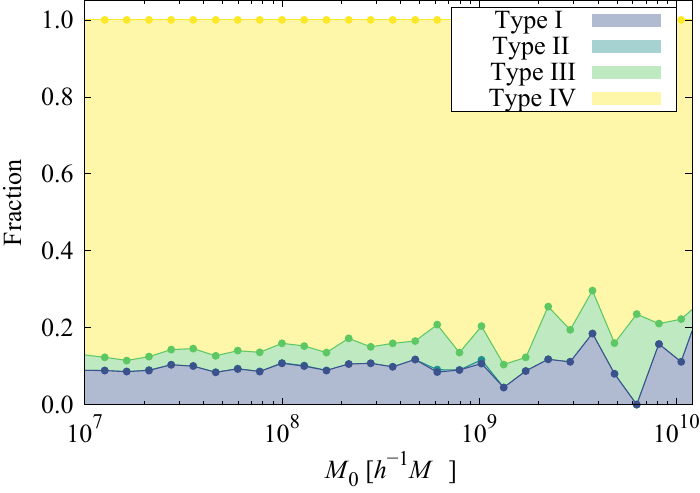}
    \end{center}
    \caption{Cumulative fraction of subhaloes with $M_0\geq 10^7\ h^{-1}M_{\odot}$, categorised into four evolutionary types, as a function of their final mass $M_0$.}
    \label{fig:type_fraction}
\end{figure}

Figure \ref{fig:type_fraction} illustrates the cumulative fraction of subhaloes with $M_0\geq 10^7\ h^{-1}M_{\odot}$ as a function of their final mass $M_0$. It is evident that Type IV is dominant, regardless of the mass of subhaloes. Type IV accounts for more than 80\%, and there is no dependence of the types on the final mass of subhaloes with $10^7\ h^{-1}M_{\odot} \leq M_0\leq 10^9\ h^{-1}M_{\odot}$. However, the fraction of subhaloes belonging to Types I and III increases at $M_0\geq 10^9\ h^{-1}M_{\odot}$ slightly.
In contrast, the late onset of the stripping phase, driven by host growth, naturally explains the absence of Type II subhaloes.
This analysis quantitatively confirms that mass loss of subhaloes due to tidal stripping by the host halo is a universal process in the mass evolutionary history of subhaloes, as discussed in the previous section.

\begin{table}
    \caption{
    Classification of mass evolutionary histories based on the two parameters, $\beta$ and $\gamma$, from equation \ref{eq:2-param} \citep[]{McBrideFakhouriMa2009}.    
    }
    \begin{center}
        \begin{tabular}{ccc}
            \hline
            Type &  Criteria &   Characteristics \\
            \hline
            I. &  $\beta <0.35$ &  exponential--like growth. \\
            II. &  $\beta -\gamma < -0.45$ & Late--time steep growth. \\
            III. & $-0.45<\beta -\gamma < 0$ & Late--time shallow growth.\\
            IV. & $\beta -\gamma >0$ & Late--time plateau.\\ 
            \hline
        \end{tabular}
    \end{center}     
    \label{tab:McBride_type}
\end{table}

\subsection{Dynamical evolution of subhaloes}

\begin{figure}[t]
    \begin{center}
      \includegraphics[width=\columnwidth]{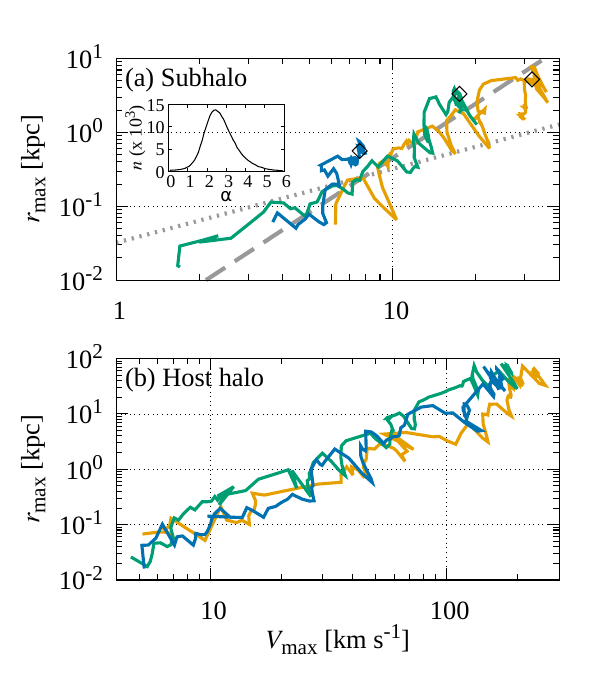}
    \end{center}
    \caption{
    The evolutionary tracks of subhaloes (\textit{upper panel}) and host haloes (\textit{lower panel}) on the $\vmax$--$\rmax$ plane. Each jagged line represents the evolutionary track of the selected haloes. In the \textit{upper panel}, the dashed line shows the $\rmax\propto\vmax^{2.45}$ relation, while the dotted line indicates the $\rmax\propto\vmax$ relation. The inset shows the distribution of $\alpha$ when each track is fitted with the function $\rmax=A\vmax^\alpha$. The open diamonds mark subhalo mass maxima.
    }
    
    \label{fig:evo-path}
\end{figure}

In this section, we present the properties of haloes on the $\vmax$--$\rmax$ plane, where $\vmax$ represents the maximum circular velocity, and $\rmax$ is the radius at which the circular velocity becomes maximum. We chose this method of presenting the data to ensure consistency with previous analyses and because $\vmax$ and $\rmax$ are easily identifiable in observational data. Additionally, we quantitatively analyse the diversity of the evolution of subhaloes on the $\vmax$--$\rmax$ plane.

Figure \ref{fig:evo-path}a shows the evolutionary tracks of subhaloes on the $\vmax$--$\rmax$ plane. 
Each jagged line represents the evolutionary track of the selected subhaloes. Seven typical subhaloes were selected, ranging from those with smaller to larger $\rmax$ values.
Since subhaloes are still small at high redshifts, their tracks begin at the lower--left corner of this plane. As haloes assemble mass from the inside out, their tracks move towards the upper--right corner. In the late stage of the accretion phase, infalling mass tends to stay in the outer regions. As the host halo grows in potential, the tidal force exerted by the host becomes stronger than the self--gravity of the subhaloes, causing them to lose mass from the outside in. Here, the tidal force primarily strips mass from the outer regions. Consequently, the subhalo tracks begin to retrace their paths, decreasing in $\rmax$ and $\vmax$. This means that the stripped--down remnants at $z=0$ resemble their high--redshift progenitors \citep{DiemandKuhlenMadau2007}.
The open diamonds represent the points at which the subhaloes reach their maximum mass. The tracks differ before (increasing in $\vmax$ and $\rmax$) and after (decreasing in both) these points. On the other hand, figure \ref{fig:evo-path}b shows the same evolutionary paths for 6 host haloes. No retreats are observed, indicating that host haloes only undergo the accretion phase.

Although figure \ref{fig:evo-path}a presents only 3 selected subhaloes, we actually investigated the evolutionary paths of all subhaloes, fitting each track with $\rmax=A\vmax^\alpha$. The inset in figure \ref{fig:evo-path}a illustrates the distribution of $\alpha$ for subhaloes, revealing a single peak, with most paths having $\alpha$ values between $1.75$ and $3.35$, and the most frequent value being $\alpha = 2.45$.
The line $\rmax\propto\vmax^{2.45}$ is shown as a grey dashed line in figure \ref{fig:evo-path}a. So far, the studies proposed that subhaloes generally evolve along the $\rmax\propto\vmax$ relation, represented by the grey dotted line \citep{DiemandKuhlenMadau2007,Hayashi+2017}. However, our results indicate that, although this trend is not significantly deviated from, individual subhaloes exhibit diverse evolutionary tracks when analysed separately.
On the other hand, figure \ref{fig:evo-path}b presents the evolutionary paths for 6 host haloes. Similar distributions of tracks are observed, where all host haloes follow the similar dynamical evolution. No retreats are evident, indicating that host haloes only undergo the accretion phase


\subsection{Formation epoch of the subhaloes and their mass distributions}
This section examines the formation time, which reflects the rate of evolution of each subhalo. Given that the mass evolution of dark matter haloes continuously changes over time, defining their formation time—which represents the timescale of halo growth—inevitably involves a certain degree of arbitrariness. As a result, several approaches to this definition have been proposed \citep[eg.,][]{Correa+2015}.
Currently, the formation redshift, $z_{\mathrm{f}}$, is often used to characterise the formation of dark matter haloes \citep{McBrideFakhouriMa2009}. The formation redshift, $z_{\mathrm{f}}$, is defined as the redshift at which the mass equals half of the final mass, $M(z_{\mathrm{f}})=M_0/2$. As we have demonstrated, the evolutionary trajectory of subhaloes is influenced not only by mass accretion but also by tidal stripping from their growing host halo. Consequently, tidal stripping is expected to have an impact on $z_{\mathrm{f}}$ as well.

\begin{figure}
    \begin{center}
        \includegraphics[width=\linewidth]{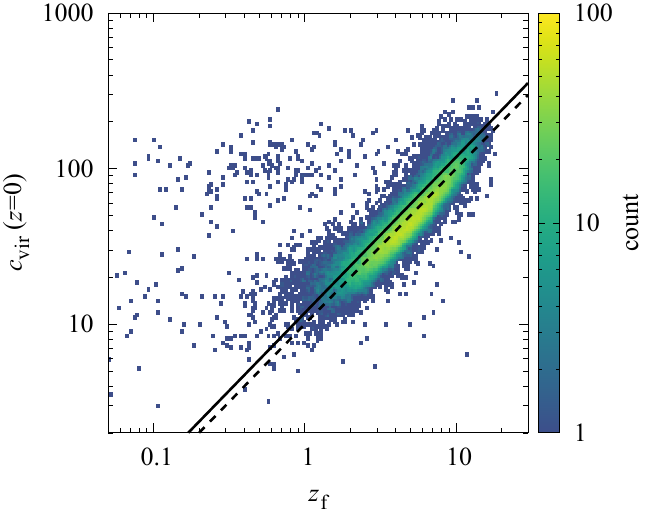}
    \end{center}
    \caption{Concentration $c_{\mathrm{vir}}$ at $z=0$ versus formation redshift $z_{\mathrm{f}}$ for all subhaloes with $M_0\geq 10^7\ h^{-1}M_{\odot}$. The colour scale indicates the number of subhaloes. The solid line represents the prediction based on \citet{Wechsler+2002}, while the dashed line represents the power--law fit ($c_{\mathrm{vir}}=9.985z_{\mathrm{f}}^{0.998}$) from this work.}
    \label{fig:conc_form-z}
\end{figure}

Figure \ref{fig:conc_form-z} shows the correlations between concentration $c_{\mathrm{vir}}$ at $z=0$ and formation redshift $z_{\mathrm{f}}$. The colour scale indicates the number of subhaloes with masses greater than $M_0=10^7\ h^{-1} M_{\odot}$.
\citet{Wechsler+2002} proposed a fitting function between concentration $c_{\mathrm{vir}}$ at $z=0$ and formation epoch $a_{\mathrm{c}}$, denoted as 
\begin{equation}
c_{\mathrm{vir}}=\frac{4.1}{a_{\mathrm{c}}} \sim 11.83z_{\mathrm{f}},
\end{equation}
for isolated dark matter haloes with $M_0>10^{12}\ h^{-1} M_{\odot}$, which is represented by the solid line in figure \ref{fig:conc_form-z}.
On the other hand, the dashed line indicates a power--law fitting curve given by
\begin{equation}
c_{\mathrm{vir}}=9.985z_{\mathrm{f}}^{0.998},
\end{equation}
to the original data. Compared to previous work on isolated massive dark matter haloes, the power--law index remains almost the same, while the intercept is smaller. 
Subhaloes with high concentrations scatter beyond the main correlation, likely due to poor fitting accuracy, and require no special attention.

This is because tidal forces effectively strip dark matter from the outer edges of subhaloes, reducing their virial radius. It can also be due to satellite galaxies losing dark matter to tidal forces from their parent halo, thereby reducing their mass, resulting in subhaloes forming at systematically higher $z_\mathrm{f}$ than the isolated subhaloes studied by \citet{Wechsler+2002}.

\subsection{Orbital parameter distribution of subhaloes and comparison with the recent observation}

From a theoretical viewpoint, it has been shown that most subhaloes undergo a tidal stripping phase. Meanwhile, recent observations have revealed the orbital motions of satellite galaxies in and around the Milky Way, enabling comparisons with theoretical simulations. Therefore, we estimate the orbital parameters of each subhalo and compare them with recent observations of satellite galaxies to determine whether the tidal forces of host galaxies have actually influenced the evolution of these satellites.
In the following analysis, we assume that subhaloes at $z=0$ behave according to Keplerian motion and identify their orbital parameters based on the method described by \citet{KhochfarBurkert2006}.

\begin{figure}[t]
    \begin{center}
        \includegraphics[width=\linewidth]{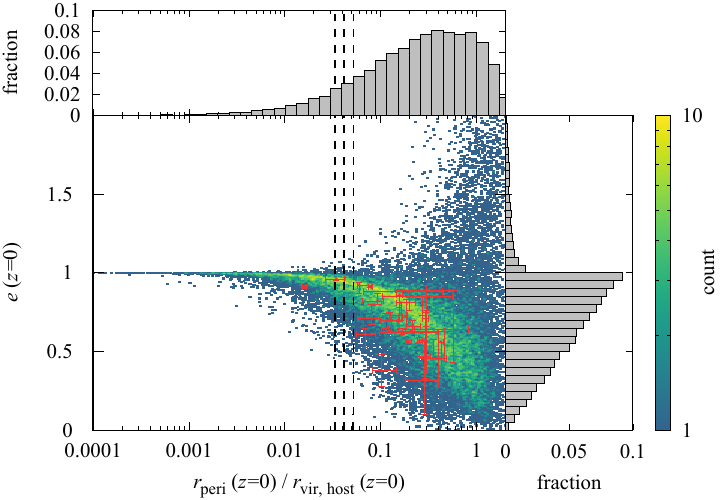}
    \end{center}
    \caption{ 
    Correlation between the orbital parameters at $z=0$ of subhaloes with $M_0\geq 10^7\ h^{-1}M_{\odot}$. The centre panel illustrates the distribution of eccentricity and pericentre distance, normalised by the virial radius of their host halo at $z=0$. The colour scale indicates the number of subhaloes. Based on the Gaia eDR 3, the filled red circles represent the median values of the orbital parameters for 31 dwarf galaxies in and around the Milky Way under the Low Mass Milky Way model \citep{Battaglia+2022}. Error bars indicate the 68\% scatter around the median. The right and top panels show histograms of eccentricity and pericentre distance, respectively. Three vertical dashed lines represent the tidal--destruction radii for dark matter subhaloes of $10^7\ h^{-1} M_{\odot}$ ($8.41$ kpc), $10^8\ h^{-1} M_{\odot}$ ($10.4$ kpc), and $10^9\ h^{-1} M_{\odot}$ ($13.0$ kpc), given by \citet{MikiMoriKawaguchi2021}
    }
    \label{fig:ecc_distribution}
\end{figure}

Figure \ref{fig:ecc_distribution} illustrates the correlation between eccentricity $e$ and pericentre distance $r_{\mathrm{peri}}$, normalised by the virial radius of their host haloes at $z=0$. This eccentricity distribution aligns broadly with the findings of \citet{KhochfarBurkert2006}. From this figure, we can discern the range of orbital parameters that enable subhaloes to withstand the tidal forces exerted by their host halo. Specifically, subhaloes with larger pericentre distances are less susceptible to these tidal forces, leading to a higher survival rate. Additionally, subhaloes with higher eccentricity experience tidal forces for shorter durations, even at the same pericentre distance, which further increases their chances of survival without being tidally destroyed.

We estimate the tidal destruction radius of satellite galaxies due to tidal stripping by solving the force balance between the self--gravity of the satellite and the tidal force by the central galaxy \citep{MikiMoriKawaguchi2021}. 
The mass model of the central galaxy is the spherical--averaged Milky Way model $M_\mathrm{MW}(r)$ given by \citet{Cautun+2020}. 
We assume an NFW sphere as a satellite galaxy with the mass $M_\mathrm{sat}$. 
Assuming 95\% of the initial mass of the satellite is tidally stripped, the tidal destruction radius $r_\mathrm{tidal}$ must satisfy the following equation,
\begin{align}
    \frac{1}{2} \bqty{\frac{G M_\mathrm{MW}\pqty{r_\mathrm{tidal} - r_{5\%}}}{\pqty{r_\mathrm{tidal} - r_{5\%}}^2} - \frac{G M_\mathrm{MW}\pqty{r_\mathrm{tidal} + r_{5\%}}}{\pqty{r_\mathrm{tidal} + r_{5\%}}^2}}
    = 0.05 \frac{G M_\mathrm{sat}}{r_{5\%}^2}
    ,
\end{align}
where $r_{5\%}$ is the radius within which 5\% of the satellite mass is contained. 
Solving this equation numerically, the resultant radius $r_\mathrm{tidal}$ is $8.41, 10.4,$ and $13.0 ~\mathrm{kpc}$ for satellite mass of $10^7, 10^8$ and $10^9 h^{-1} M_\odot$, respectively. 
The dashed lines in figure \ref{fig:ecc_distribution} represent the tidal destruction radii at $10^7$, $10^8$, and $10^9 h^{-1} M_{\odot}$. 

For comparison, we have overplotted the orbital parameters of satellites in and around the Milky Way, as derived from a high--precision orbital analysis by \citet{Battaglia+2022} using data from Gaia’s Early Data Release 3 (eDR3). \citet{Battaglia+2022} analysed the spatial distribution and the distribution within the colour--magnitude and intrinsic motion planes of individual objects with complete Gaia eDR3 positional measurements, focusing on 74 Local Group dwarf galaxies. They determined the intrinsic motion of these entire systems, including 14 galaxies located beyond the virial radius of the Milky Way, up to approximately $430$ kpc away. 
As shown in figure \ref{fig:ecc_distribution}, these data, including the error bars, are consistent with the \textsc{Phi--4096} results, demonstrating no significant discrepancy between the CDM model and the actual orbital motions of satellite galaxies in and around the Milky Way. Notably, both theory and observation reveal a sharp decrease in the number of satellite galaxies within the destruction radius. Only satellite galaxies with greater eccentricities survive inside this radius, which strongly reflects the characteristics of the tidal stripping process.  This agreement leads us to the intriguing conclusion that most of the observed satellite galaxies in the Milky Way have undergone their dynamical and mass evolution under the influence of tidal forces, resulting in their current state.

\section{Summary and Discussion}

In this study, the mass evolutionary histories of dark matter subhaloes were examined in detail using the ultra--high--resolution cosmological simulation \textsc{Phi--4096} within the framework of the CDM model. The mass evolution of the most massive progenitors of subhaloes was normalised by their virial mass at redshift $z=0$ and analysed as a function of redshift. The results reveal a distinct pattern where subhaloes experience an initial phase of mass accretion, followed by a phase of mass loss due to tidal stripping from growing their host halo. This two--phase evolution is evident across various mass ranges, with the peak redshift, where the median mass reaches its maximum, decreasing as the final subhalo mass increases. 

The study also categorised the dynamical evolutionary histories of subhaloes into four types, based on parameters defined by \citet{McBrideFakhouriMa2009}. The analysis shows that Type IV subhaloes, characterised by a plateau in their late--time evolution, dominate across all mass ranges, particularly in less massive subhaloes. This classification highlights the diversity in the evolutionary paths of subhaloes, with a significant portion undergoing considerable mass loss due to tidal stripping.
Further, this study investigated the formation epoch of subhaloes, defining the formation redshift $z_{\mathrm{f}}$ as the point where the subhalo mass equals half of its final mass. The correlation between the concentration parameter $c_{\mathrm{vir}}$ at $z=0$ and $z_{\mathrm{f}}$ was analysed, showing a relationship consistent with previous studies but with slight deviations due to the effects of tidal stripping. This relationship provides insights into the formation and subsequent evolution of subhaloes within their host environments.
Finally, the orbital parameters of subhaloes were compared with recent observational data from the Gaia satellite. The eccentricity and pericentre distances of subhaloes were analysed, showing broad consistency with the orbital motions of the satellites in the Milky Way. This comparison supports the validity of the CDM model in explaining the dynamical behaviour of subhaloes, as no significant discrepancies were found between the simulation results and the observed data.

Here, we examine the numerical convergence of this study, following the two conditions proposed by \citet{vandenBoschOgiya2018}, who discuss numerical artefacts in tidal evolution. The first condition, related to force softening, is: 
\begin{equation}
\frac{r_\mathrm{vir}}{\varepsilon} > \frac{0.73 \,c_\mathrm{vir}^{1.63}}{\ln(1 + c_\mathrm{vir}) - c_\mathrm{vir}/(1 + c_\mathrm{vir})},
\end{equation}
where $\varepsilon (= 0.06 \, h^{-1} \,\mathrm{kpc})$ is the softening parameter. The original formula uses the half--mass radius, but we have converted it to the virial radius $r_\mathrm{vir}$ here.
Although the latest $c-M$ relation takes a different form as examined by \citet{KanedaMoriOtaki2024}, if we here adopt the $c-M$ relation they used, $c_\mathrm{vir} = 5.26 (M_\mathrm{vir}/10^{14} h^{-1} M_\odot)^{-0.1}$ given by \citet{Neto+2007},
this criterion can be rewritten as $M_\mathrm{vir} > 1.1 \times 10^7 \ h^{-1} M_\odot$ with simple manipulation and numerical calculation.
The second condition concerns discreteness noise, requiring $N_\mathrm{p} > 250$ particles, corresponding to $M_\mathrm{vir} > 1.3 \times 10^6\ h^{-1} M_\odot$. 
Taking these two conditions into account, the minimum mass limit of $M_\mathrm{vir} \ge 10^7\ h^{-1} M_\odot$ adopted in this study is reasonable and ensures robust results.

The CDM model has been successful in explaining the large--scale structure of galaxy distributions, as noted above. However, it faces significant difficulties at subgalactic scales. To address these issues, alternative dark matter models, such as warm dark matter, fuzzy dark matter, and self--interacting dark matter, have been proposed. Nevertheless, observational constraints, particularly from galactic satellites, have placed significant limits on the parameters of these models \citep{Hayashi+2021a, Nadler+2021, DalalKravtsov2022, Dekker+2022}.
In this study, we analysed the mass evolution of dark matter haloes within the framework of the CDM model and compared our results with observations from the Gaia satellite. Gaia provides an exceptional dataset for testing the CDM model at subgalactic scales, thanks to its precise measurements of stellar positions, velocities, and distances. Future studies are expected to compare subhalo mass evolution and orbital elements within these alternative dark matter models using the latest data. Such comparisons will not only clarify the relevance of these models but also refine their parameters, thereby enhancing our understanding of the role of dark matter, including the possibility of multiple components of dark matter particles, in the formation of cosmic structure.


\begin{ack}
We thank the anonymous reviewer for the comments, Tomoaki Ishiyama for providing the results of the cosmological $N$--body simulation ‘Phi-4096’ and Giuseppina Battaglia for the orbital parameters of the Milky Way satellites. We received valuable comments from Kohei Hayashi, Yohei Miki, Takanobu Kirihara, and Andrea Ferrara.
This work was supported by JSPS KAKENHI Grant Numbers JP23KJ0280, JP24K07085 and JP24K00669.
Numerical computations were performed with computational resources provided by the Multidisciplinary Cooperative Research Program in the centre for Computational Sciences, University of Tsukuba.
\end{ack}


\bibliographystyle{apj}
\bibliography{kazuno+2024}


\end{document}